\documentclass[a4paper,fleqn]{cas-dc}
\usepackage[numbers]{natbib}
\usepackage{longtable}
\begin{document}
\let\WriteBookmarks\relax
\def\floatpagepagefraction{1}
\def\textpagefraction{.001}

\shorttitle{Luminescence from oxygen vacancies in Lu$_{2}$SiO$_{5}$ crystal and ceramics at room temperature}    

\shortauthors{M. V. Belov, S. A. Koutovoi, V. A. Kozlov et al}  

\title [mode = title]{Luminescence from oxygen vacancies in Lu$_{2}$SiO$_{5}$ crystal and ceramics at room temperature}  

\author[1]{M.~V.~Belov}
\ead{belovmv@lebedev.ru} 
\affiliation[1]{organization={P.N. Lebedev Physical Institute of the Russian Academy of Sciences},
            addressline={Leninskii prospekt 53}, 
%            city={Moscow},
%          citysep={}, % Uncomment if no comma needed between city and postcode
            postcode={119991}, 
            state={Moscow},
            country={Russia}}
            
\author[2]{S.~A.~Koutovoi}
\ead{koutov@lsk.gpi.ru} 
\affiliation[2]{organization={A.M. Prokhorov General Physics Institute of the Russian Academy of Sciences},
            addressline={Vavilova str. 38}, 
%            city={Moscow},
%          citysep={}, % Uncomment if no comma needed between city and postcode
            postcode={119991}, 
            state={Moscow},
            country={Russia}}

\author[1]{V.~A.~Kozlov}
\ead{kozlov@lebedev.ru}

\author[1]{N.~V.~Pestovskii}
\cormark[1]
\cortext[1]{Corresponding author}
\ead{pestovskii@lebedev.ru}
\ead{pestovsky@phystech.edu}

\author[1]{S.~Yu.~Savinov}
\ead{savinov@lebedev.ru}

\author[2]{A.~I.~Zagumennyi}
\ead{zagumen@lsk.gpi.ru}  

\author[2]{Yu.~D.~Zavartsev}
\ead{zavart@lsk.gpi.ru}   

\author[1]{M.~V.~Zavertyaev}
\ead{zavert@lebedev.ru }
            
            %\cormark[1]
%\cortext[1]{Corresponding author}
%\ead{pestovskii@lebedev.ru}
%\ead{pestovsky@phystech.edu}

%\author{S. Yu. Savinov}
%\ead{savinov@lebedev.ru}

\begin{abstract}
Photoluminescence (PL) of Lu$_{2}$SiO$_{5}$ crystal and ceramics with a high concentration of oxygen vacancies (about ~0.5 at.\%) is studied. Oxygen vacancies were created using two ways. The first method is a growth of crystal from the non-stoichiometric Lu$_{2}$Si$_{0.98}$O$_{4.96}$ melt and the second one is a doping of Lu$_{2}$SiO$_{5}$ matrix with divalent Ca$^{2+}$ ions at the concentration of 1 at.\%. For the first time we observe a fairly bright room-temperature ultraviolet PL of both crystal grown from the Lu$_{2}$Si$_{0.98}$O$_{4.96}$ melt and Lu$_{2}$SiO$_{5}$ ceramics doped with Ca$^{2+}$ ions. A band at 290~nm in the PL spectrum of the crystal and a band at 283~nm in the PL spectrum of the calcium-doped ceramics are observed. The spectral and kinetic properties of these bands are
close to each other. This fact indicates similar origins of these bands. The results of the work show that the studied ultraviolet luminescence is related to oxygen vacancies in lutetium oxyorthosilicate
\end{abstract}


\begin{highlights}
\item Bright luminescence from oxygen-deficient Lu$_{2}$SiO$_{5}$ at room-temperature is observed.
\item Emission at 280-290~nm is dipole-resolved and due to oxygen vacancies.
\item Parameters of the emission are close to STE luminescence from Lu$_{2}$SiO$_{5}$.

\end{highlights}

% Keywords

\maketitle

% Main text
\section{Introduction}\label{Introduction}

Luminescence of non-doped crystalline matrix (luminescence of defects, excitons etc) contains information about an evolution of electronic excitations, non-radiative decay and energy transfer channels. Lutetium oxyorthosilicate crystal Lu$_{2}$SiO$_{5}$ (LSO) activated with Ce$^{3+}$ ions is well-known as a highly efficient radiation-hard scintillator~\cite{1,2,3,4,5,6,7}. This crystal is widely
used in many practical applications, particularly in positron-emission tomography, high-energy physics etc.

Intrinsic luminescence of LSO matrix without additional doping was studied in Refs.~\cite{8,9,10}. It was found that at liquid helium temperature, luminescence of pure LSO is composed of two emission bands at $\sim$250 and $\sim$360~nm. It is shown~\cite{8,9,10} that this emission is related to the radiative decay of self-trapped excitons (STE). At room temperature, LSO intrinsic luminescence has not been yet observed. Determination of the intrinsic luminescence mechanisms is required for revealing the scintillation mechanism in LSO doped with Ce$^{3+}$.
Indeed, at least about $\sim$50\% of emitting Ce$^{3+}$ ions in LSO at room temperature are excited due to STE destruction~\cite{4}.

In Ref.~\cite{2} it was for the first time found, that an additional doping of LSO:Ce with Ca$^{2+}$ or Mg$^{2+}$ ions leads to a significant improvement in scintillation and growth properties. For this reason, the LSO:Ce:Ca composition and its modifications now are the most used in applications among scintillators based on LSO~\cite{3,6,7}. The physical mechanism of improvement in scintillation and growth properties induced by calcium co-doping (an increase in the light yield, shortening in the scintillation decay time and decreasing in a tendency of an LSO crystal to crack during the growth and cutting into detector pixels) is now extensively studied~\cite{11,12,13,14,15,16,17,18,19,20}. However, the origins of these phenomena are now still unclear.

Numerical calculations prove~\cite{18,20} that doping of LSO crystalline lattice with divalent Ca$^{2+}$ ions leads to substitution of trivalent lutetium Lu$^{3+}$ ions by Ca$^{2+}$ ions. As a result, a $\{$Ca$_{\mathrm{Lu}}$+V$_{\mathrm{O}}\}$ defect consisting of Ca$^{2+}$ ion and oxygen vacancy V$_{\mathrm{O}}$ is created. A complexity of LSO:Ca luminescence mechanisms is seen in the following example. At first sight, an increase in the defect concentration caused by doping with Ca$^{2+}$ ions should lead to increase in the concentration of electron traps. Thereby, an increase in the Ca$^{2+}$ concentration in LSO should  deteriorate its scintillation parameters. However, it is observed experimentally~\cite{12} that the concentration of these traps is decreased with an increase in the Ca$^{2+}$ concentration. This fact is not explained now unambiguously.

Until now, there were no studies of luminescence from pure LSO matrix not containing cerium but doped with Ca$^{2+}$ ions. Also, up to now, there were no investigations of luminescence from pure LSO matrix grown from a non-stoichiometric melt with oxygen deficiency. However, these studies are required to reveal mechanisms of evolution of electronic excitations  in LSO:Ca and LSO:Ca:Ce. Indeed, luminescence studies of LSO:Ca can provide an information about defects created in LSO due to doping, especially the lifetime of its electronic
excitation and energetic characteristics. In general, it is now unclear if the doping with Ca$^{2+}$  ions leads to formation of oxygen vacancies only in LSO structure or induces some additional processes. A comparison of luminescence characteristics for LSO:Ca and pure LSO grown from a non-stoichiometric melt with oxygen deficiency can provide an answer on this question.

 The purpose of this study is to study the spectral and kinetic properties of luminescence from LSO with the oxygen vacancies concentration of $\sim$0.5~at.\% and to investigate an origin of this luminescence. Creation of additional oxygen vacancies in studied samples was made using two ways. First one is a Czochralski growth of LSO crystal from a non-stoichiometric melt and second one is a synthesis of LSO ceramics doped with Ca$^{2+}$ ions.
 
\begin{table*}
\caption{\label{tab:table3}Characteristics of studied samples}
\begin{tabular}{llll}
Chemical formula&Type&Designation in this work&Dopants\\
\hline
Lu$_{2}$Si$_{0.98}$O$_{4.96}$ (composition of melt) &crystal&&none\\
Lu$_{2}$SiO$_{5}$&ceramics&LSO&none\\
Ca$_{0.08}$Lu$_{1.92}$SiO$_{4.96}$&ceramics&LSO:Ca$^{2+}$&1.0~at.\% Ca$^{2+}$\\
Ca$_{0.08}$Lu$_{1.92}$Si$_{0.947}$P$_{0.058}$O$_{4.999}$&ceramics&LSO:Ca$^{2+}$:P$^{5+}$&1.0~at.\% Ca$^{2+}$ and 0.7~at.\%~P$^{5+}$\\
\end{tabular}
\end{table*}

\section{Experimental detalis}\label{Experimental detalis}
\subsection{LSO samples}

Characteristics of the samples are summarized in Table~1.
Studied LSO samples were synthesized from Lu$_{2}$O$_{3}$, SiO$_{2}$, CaO and P$_{2}$O$_{5}$ powders with the mass fractions of primary substances at least 99.99~wt.\%. The LSO crystal was grown from non-stoichiometric melt Lu$_{2}$Si$_{0.98}$O$_{4.96}$ using Czochralski technique from an iridium crucible
with the diameter of 40~mm heated by radiofrequency electromagnetic field in the atmosphere of 99.9\% Ar + 0.1\%O$_{2}$. The growth speed was of 3~mm/h and the rotational speed was of 10~rpm. The composition of the melt was Lu$_{2}$Si$_{0.98}$O$_{4.96}$ where there was a deficiency in oxygen and silicon with respect to lutetium and an excess of oxygen with respect to silicon. The oxygen vacancy concentration induced by the non-stoichiometry of the melt was of 0.5~at.\%.

The studied sample was made from the bottom part of the grown boule because the intensity of luminescence observed in this work was the highest in the bottom part of the boule. The top of the boule was colorless, but the bottom of the boule had a weak green color caused by higher defect concentration. The studied LSO crystal was contaminated with small impurities of cerium and other elements. The concentrations of impurities in the sample were measured using
Glow discharge mass spectrometry (GDMS) technique. It is found that the mass fractions of the main impurities in the crystal were Na - 0.75$\cdot$10$^{-4}$~wt.\%, Mg - 0.40$\cdot$10$^{-4}$~wt.\%, Al - 1.5$\cdot$10$^{-4}$~wt.\%, Cl - 8.0$\cdot$10$^{-4}$~wt.\%, Ca – 5.4$\cdot$10$^{-4}$~wt.\%, Cr - 1.0$\cdot$10$^{-4}$~wt.\%, Fe - 1$\cdot$10$^{-4}$~wt.\%, Y - 10$\cdot$10$^{-4}$~wt.\%, Ce - 0.5$\cdot$10$^{-4}$~wt.\%, Gd - 0.25$\cdot$10$^{-4}$~wt.\%, Yb - 2.0$\cdot$10$^{-4}$~wt.\% and Ir - 7.3$\cdot$10$^{-4}$~wt.\%. Other elements had the total mass fraction less than 10$^{-4}$~wt.\%. Thereby, the total impurity content in
the sample was less than 4$\cdot$10$^{-3}$~wt.\%.

Ceramic samples Lu$_{2}$SiO$_{5}$ (LSO), Ca$_{0.08}$Lu$_{1.92}$SiO$_{4.96}$ (LSO:Ca$^{2+}$) and Ca$_{0.08}$Lu$_{1.92}$Si$_{0.947}$P$_{0.058}$O$_{4.999}$ (LSO:Ca$^{2+}$:P$^{5+}$) were created in air at atmospheric pressure using
a two-stage high-temperature synthesis of mechanical mixture of oxides. In order to obtain a homogeneous distribution of Ca$^{2+}$ ions in the mixture we dissolved the CaO precursor in HNO$_{3}$ acid. Then we jumbled the mixture in an aqueous solution containing Ca$^{2+}$ ions. In order to alloy the ceramics with phosphorus we dissolved P$_{2}$O$_{5}$ precursor in distilled water and jumbled
the mixtures of oxides in aqueous solution containing Ca$^{2+}$ and P$^{5+}$ ions.

At the first stage of the synthesis, the mechanical mixture was annealed at the
temperature of 1100$^{\circ}$C in atmospheric air. After that, these tablets were annealed again at the temperature of 1600$^{\circ}$C in atmospheric air (the second stage of the synthesis). As a result of the solid phase synthesis, the polycrystalline samples were created with known chemical composition. It should be especially noted that all the LSO ceramic samples studied in this work did not contain cerium impurities at measurable concentrations. Indeed, all the precursors
did not contain cerium at appraisable levels. Also, cerium ions could not contaminate the ceramic samples during the synthesis.

As it was mentioned above, according to Refs.~\cite{18,20}, doping of LSO with Ca$^{2+}$ ions lead to creation of $\{$Ca$_{\mathrm{Lu}}$+V$_{\mathrm{O}}\}$ defects including oxygen vacancies. The main reason of the oxygen vacancy creation is the compensation of the charge imbalance. Indeed, a substitution of a Lu$^{3+}$ ion by a Ca$^{2+}$ ion lead to formation of excess negative charge. On the other hand, it is well-known that oxygen ions in LSO are negatively charged. Thereby, formation of oxygen vacancy near the Ca$^{2+}$ ion leads to compensation of the charge imbalance.

According to the above argumentation, simultaneous doping of LSO with Ca$^{2+}$ and P$^{5+}$ in equal concentration should not lead to oxygen vacancy formation. Indeed, in this case, the negative charge imbalance induced by Ca$^{2+}$ ions substituting Lu$^{3+}$ is equilibrated by the positive charge provided by P$^{
5+}$ ions substituting Si$^{4+}$ ions. In general, a substitution of Si$^{4+}$ ions by P$^{5+}$ ions is well-known~\cite{21,22,23,24,25,26}. An experimental proof of the incorporation of phosphorus into the glassy SiO$_{2}$ in the pentavalent state is obtained, particularly, in Ref.~\cite{22}. It should be noted
that glassy SiO$_{2}$ is composed of the SiO$_{4}$ tetrahedra which are also the building blocks of LSO~\cite{20}.

Thereby, based on the given data we can conclude that the LSO:Ca$^{2+}$:P$^{5+}$ sample simultaneously doped with Ca$^{2+}$ and P$^{5+}$ ions contains a significantly smaller concentration of oxygen vacancies than the LSO:Ca$^{2+}$ sample. However, this sample contains calcium. Consequently, based on a
comparison of luminescence parameters for LSO:Ca$^{2+}$ and LSO:Ca$^{2+}$:P$^{5+}$ samples we can reveal separately a role of oxygen vacancies and calcium doping in LSO luminescence.

It should be noted that the real concentrations of oxygen vacancies in all studied samples were not measured directly. We estimated their concentrations from the relations between the weights of the initial materials. Thereby, the vacancy concentrations can be estimated from the chemical formulas of the samples. We suppose that the oxygen vacancy concentration in the non-stoichiometric crystal grown from the Lu$_{2}$Si$_{0.98}$O$_{4.96}$ melt should be lower than 0.5~at.\% (the oxygen vacancy concentration in the melt) because the growth of nearly stoichiometric single crystal from liquid melt is more energetically favorable. Simultaneously, a synthesis of the ceramics provide a possibility to create more defective material than the growth from a liquid melt. Thereby, we suppose that the vacancy concentrations in the ceramics samples are higher than in the studied crystal and they are close to the concentrations estimated from the chemical formulas.

\subsection{Optical measurements}

Photoluminescence (PL) of LSO samples was excited at 205~nm (6.05~eV). Exciting
radiation was generated by a Sol Intstruments CF-131 sapphire:Ti$^{2+}$ laser operated at forth harmonic analogous to Ref.~\cite{27}. An optical pumping of the laser was carried out using the radiation of a Sol Instruments LF-117 YAG:Nd$^{3+}$ laser 532~nm. The duration of exciting radiation was of $\sim$7~ns, the energy of each pulse was of $\sim$5~$\mathrm{\mu}$J. The frequency of laser pulses was of 10~Hz. In order to suppress the radiation of the first, second and third laser harmonics at 820, 410 and 273~nm, the exciting radiation was transmitted firstly through a Pellin–Broca prism and then through a 60$^{\circ}$-prism. All the prisms were made of fused quartz. As a result of filtering, parasitic radiation was completely suppressed.

Measurements of the PL spectral and kinetic characteristics were carried out analogous to Refs.~\cite{27,28,29}. PL emission was collected in standard 90$^{\circ}$-geometry analogous to~\cite{27,28} using quartz optical fiber with the diameter of 1~mm. PL energy spectral density (spectrum) the emission was studied using an OCEAN FLAME-S-XR1-ES spectrometer with the entrance slit width of 50~$\mathrm{\mu}$m. The spectral resolution of the spectrometer was of 2~nm. The spectral region of the spectrometer was of 200-1024~nm. The spectrometer accumulated the studied emission during
65~s. Thereby, spectra in this work correspond to the sum of the energy spectral densities of 650~pulses. All spectra in this work were corrected according to the spectral sensitivity curve of the measuring system. Correction was made using the etalon halogen and deuterium lamps.

Temporal dependence of the PL intensity (kinetics) at various wavelengths was
measured using the following apparatus. PL was directed into the input slit of a Solar TII MS2004 monochromator with the diffraction grating (1200~gr/mm, the inverse linear dispersion is of 4~nm/mm). The intensity of the light signal was measured using a Hamamatsu H3695-10 photomultiplier (PMT) mounted to the output slit of the monochromator. The PMT photocurrent was measured by a Tektronix MSO6-2500BW oscilloscope with a bandwidth of 2.5~GHz. The total time resolution of the system was $\sim$1~ns. All kinetic curves in this work are obtained by the averaging of 100 curves corresponding to each PL pulse. All studies were made at room temperature in atmospheric air.

\section{Results and discussion}\label{Results and discussion}

\begin{figure}
\includegraphics{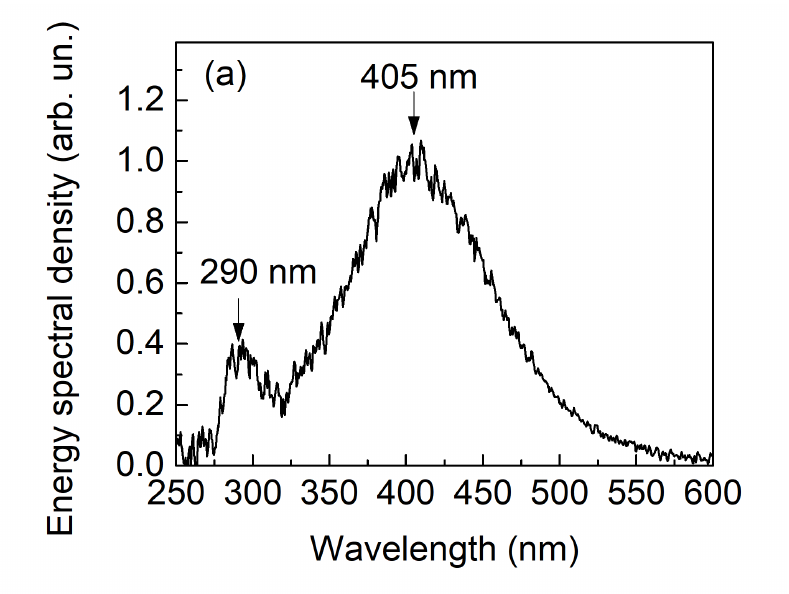}
\includegraphics{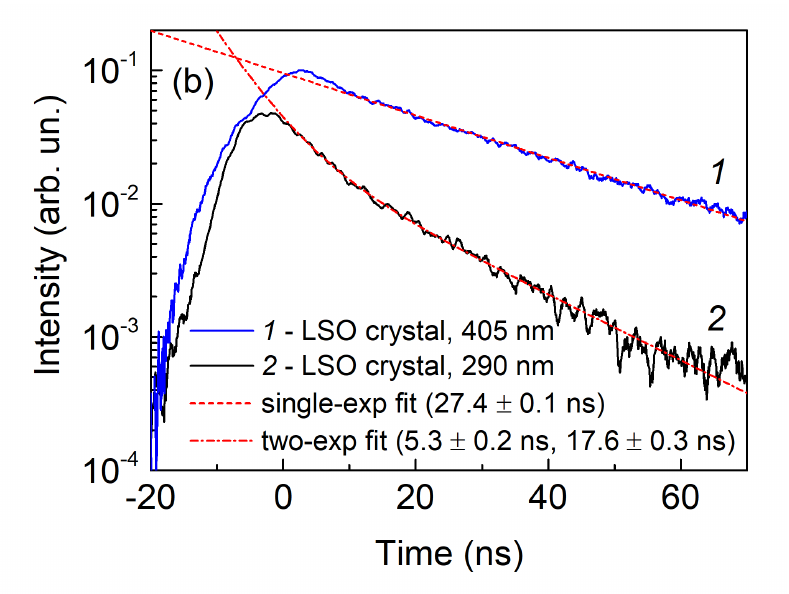}
\caption{\label{fig:epsart}Fig. 1. (a) – PL spectrum of the crystal grown from the Lu$_{2}$Si$_{0.98}$O$_{4.96}$ melt excited at 205~nm; (b) – PL kinetics of this crystal at 405~nm (1) and 290~nm (2).}
\end{figure}

PL spectrum of the crystal grown from the Lu$_{2}$Si$_{0.98}$O$_{4.96}$ melt is presented in Fig.~1~(a). It is seen that the spectrum consists of two wide bands with the maxima at 290 and 405~nm. The PL kinetics of these bands is presented in Fig.~1~(b). PL kinetics at 405~nm is adequately described by a single-exponent curve:

\begin{equation}
	I(t) = a \exp \left( -t/\tau \right),
\end{equation}

where $a$ and $\tau$ are the fitting parameters. Fitting of the curve in Fig.~1~(b,2) corresponding to PL at 405~nm gives $\tau$~=~(27.4~$\pm$~0.1)~ns. Kinetics of PL at 290~nm is described by a two-exponent function:

\begin{equation}
	I(t) = a_{1} \exp \left( -t/\tau_{1} \right) + a_{2} \exp \left( -t/\tau_{2} \right),
\end{equation}

where $a_{1}$, $a_{2}$, $\tau_{1}$ and $\tau_{2}$ are the fitting parameters. For the PL kinetics at 290~nm we obtain $\tau_{1}$ = (5.3~$\pm$~0.2)~ns, $a_{1}$~=~(25~$\pm$~2)\%, $\tau_{2}$~=~(17.6~$\pm$~0.3)~ns and $a_{2}$~=~(75~$\pm$~2)\%. It is well-known that the room-temperature emission of Ce$^{3+}$ in LSO is at 405~nm~\cite{28}. At high Ce$^{3+}$ concentrations ($\sim$1~at.\%), the decay time of the Ce$^{3+}$ emission is about 35-40~ns~\cite{1,2,3,28}. This quantity is close to the decay time at 405~nm observed in PL of the non-stoichiometric crystal (27.4~$\pm$~0.1~ns).
Thereby, we can make an assumption that the PL band at 405~nm in spectrum of the crystal grown from the Lu$_{2}$Si$_{0.98}$O$_{4.96}$ melt is related to Ce$^{3+}$
-emission. Cerium contaminated the crystal because the used iridium crucible was used previously for growing crystals activated by cerium.

The PL band at 290~nm in spectrum of LSO is observed for the first time. Further studies in this work are devoted to investigate its nature. As a hypothesis, let us assume that the emission at 290~nm is due to optical transitions in oxygen vacancies. It is well-known that these defects lead to a bright luminescence from SiO$_{2}$~\cite{30,31} and other substances.

In order to verify this hypothesis, we synthesized samples of LSO ceramics in which oxygen vacancies were artificially created by doping of LSO with Ca$^{2+}$ ions. As it was mentioned in Sec. 2.1, Ca$^{2+}$ ions substitutes Lu$^{3+}$ ions~\cite{18,20}. Charge imbalance is equilibrated due to rejection of oxygen atoms from crystal. Each oxygen atom in LSO have the electric charge of 2$^{-}$. Thereby, one oxygen vacancy is formed per two introduced Ca$^{2+}$ ions. The concentration of doping impurities and oxygen vacancies in ceramic samples were determined from the chemical formulas of the samples. We assumed that 100\% is a total number of ions in LSO.

Thereby, 1.0~at.\% of Ca$^{2+}$ ions creates 0.5~at.\% of oxygen vacancies. On the other hand, analogous to SiO$_{2}$~\cite{22} doping with P$^{5+}$ ions leads to substitution of Si$^{4+}$ by them in LSO structure and induces the charge imbalance of the opposite sign. Indeed, simultaneous doping with Ca$^{2+}$ and P$^{5+}$ ions occurs with a charge compensation according to the scheme

\begin{equation}
	\mathrm{Lu}^{3+} + \mathrm{Si}^{4+} \to \mathrm{Ca}^{2+} + \mathrm{P}^{5+}.
\end{equation}

In this case, the following equality of electric charges is satisfied: (3$^{+}$)~+~(4$^{+}$)~=~(2$^{+}$)~+~(5$^{+}$). Consequently, simultaneous doping with Ca$^{2+}$ and P$^{5+}$ in equal concentrations do not lead to formation of additional oxygen vacancies in LSO in contrast to doping with Ca$^{2+}$ ions only.

\begin{figure}
\includegraphics{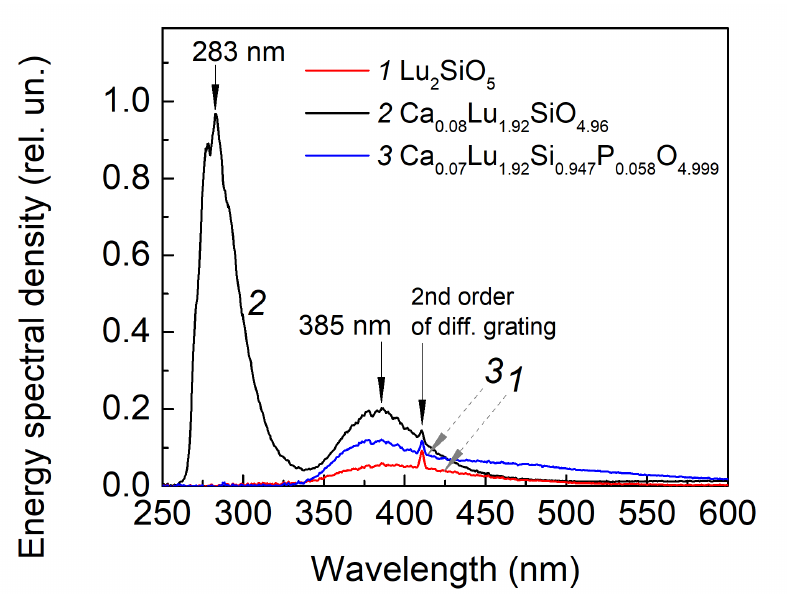}
\caption{\label{fig:epsart} PL spectra of LSO ceramics excited at 205~nm: 1-stoichiometric LSO, 2-LSO:Ca$^{2+}$, 3-LSO:Ca$^{2+}$:P$^{5+}$.}
\end{figure}

In Fig.~2, a comparison of the PL spectra for stoichiometric LSO ceramics without
additional doping, LSO:Ca$^{2+}$ ceramics doped with 1.0~at.\% Ca$^{2+}$ ions and LSO:Ca$^{2+}$:P$^{5+}$ ceramics simultaneously doped with 1.0~at.\% Ca$^{2+}$ and 0.7~at.\%~P$^{5+}$ ions is presented. It is seen that the phosphorus and calcium concentration in this composition matches with the accuracy of $\sim$30\%. It should be noted that the precursors of ceramics did not contain cerium at measurable concentrations. Also, the method of synthesis of the LSO ceramic samples prevented contamination of the samples with cerium ions at its all stages.

It is seen from Fig.~2 that the non-doped stoichiometric LSO ceramics has a weak PL band at 385~nm. Doping of LSO ceramics with Ca$^{2+}$ creating the 0.5~at.\% of oxygen vacancies leads to the increase in the PL energy spectral density at 385~nm by $\sim$4 times. Also, doping with Ca$^{2+}$ leads to formation of a new PL band at 283~nm. Its energy spectral density is $\sim$20~times higher compared with the band at 385~nm in the PL spectrum of the stoichiometric ceramic sample.

Simultaneous doping of LSO with Ca$^{2+}$ and P$^{5+}$ ions in similar concentrations which is not accompanied by formation of oxygen vacancies leads to complete suppression of this PL band at 283~nm. However, the bad at 385~nm persists and its PL energy spectral density is of $\sim$2 times higher than in the PL spectrum of the non-doped LSO ceramics and thereby of $\sim$2 times weaker than in the PL spectrum of LSO:Ca ceramics.

\begin{figure}
\includegraphics{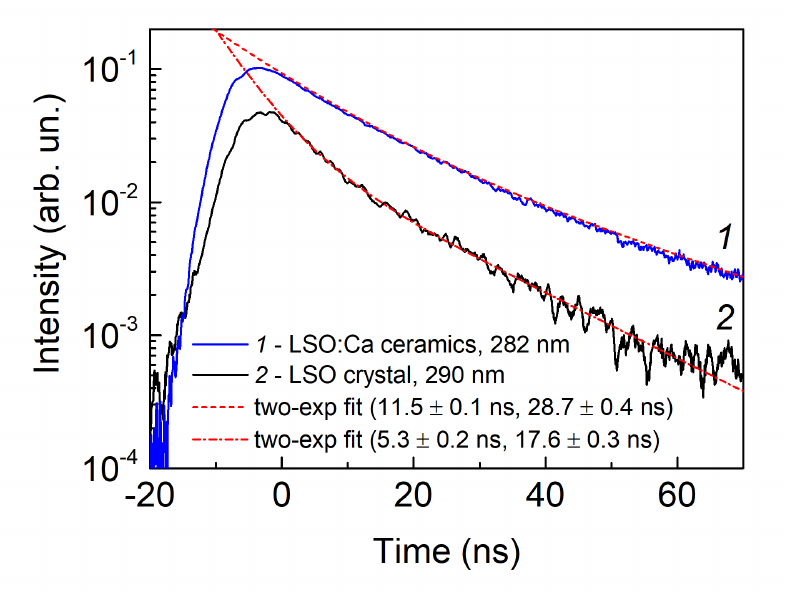}
\caption{\label{fig:epsart} PL kinetics of a Ca$_{0.08}$Lu$_{1.92}$SiO$_{4.96}$ ceramics at 282~nm~(1) and a crystal grown from the Lu$_{2}$Si$_{0.98}$O$_{4.96}$ melt at 290~nm~(2). Excitation wavelength in both cases is at 205~nm.}
\end{figure}

The results presented in Fig.~2 show that a creation of $\sim$0.5~at.\% oxygen vacancies in LSO ceramics leads to fairly bright luminescence at $\sim$280~nm. The wavelength of this emission is close to the position of the PL band of LSO crystal grown from the non-stoichiometric melt (Fig.~1). A comparison of the PL kinetics of the LSO:Ca$^{2+}$ ceramics and the crystal grown from the Lu$_{2}$Si$_{0.98}$O$_{4.96}$ melt is presented in Fig.~3.

It is seen from Fig.~3, that the kinetic parameters of the bands are close to each other and they have only slight difference. Fitting of the kinetics of LSO:Ca$^{2+}$ at 283~nm by Eq.~(2) gives $\tau_{1}$~=~(11.5~$\pm$~0.1)~ns, $a_{1}$~=~(53~$\pm$~1)\%, $\tau_{2}$~=~(28.7~$\pm$~0.4)~ns and $a_{2}$ = (47~$\pm$~1)\%. Thereby, the ceramics and crystal PL in the region of 280-290~nm is characterized by close decay times. Consequently, a similarity in the kinetic and spectral parameters of these ultraviolet bands of the LSO:Ca$^{2+}$ ceramics and the crystal grown from the non-stoichiometric melt Lu$_{2}$Si$_{0.98}$O$_{4.96}$
makes it possible to conclude that the origins of these bands are similar. This conclusion is supported by the coincidence in the chemical compositions of the materials and in the measuring conditions of the PL parameters.

Consequently, according to the above analysis, these bands both in the ceramics
LSO:Ca$^{2+}$ and in the non-stoichiometric LSO crystal are related to oxygen vacancies. The nanosecond duration of the emission shows that the corresponding radiative transitions are dipole-resolved. We suppose that an analogous luminescence bands there are in spectrum of glassy SiO$_{2}$ at 280~nm. These bands are emitted during the transitions between the states of oxygen vacancies without change in the spin of states. It is well-known that the duration of this
emission is of a few nanoseconds~\cite{30,31}.

\begin{figure}
\includegraphics{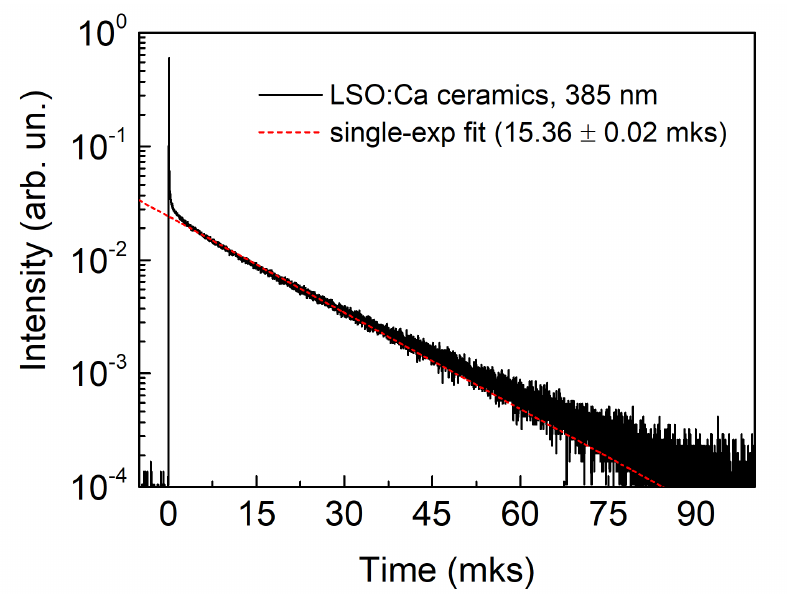}
\caption{\label{fig:epsart} The PL kinetics of the Ca$_{0.08}$Lu$_{1.92}$SiO$_{4.96}$ ceramics at 385~nm excited at 205~nm.}
\end{figure}

The kinetic of the PL band at 385~nm for the 
LSO:Ca$^{2+}$
%Ca$_{0.08}$Lu$_{1.92}$SiO$_{4.96}$ 
ceramics is presented in Fig.~4. A fitting of the curve by Eq.~(1) gives the PL decay time $\tau$~=~(15.36~$\pm$~0.2)~$\mathrm{\mu}$s. This time is 10$^{3}$-10$^{4}$ times longer than the decay times of the bands in the region of 280-290~nm. It is obvious that the origin of the band at 385~nm differs from the band origin of PL at 405~nm in the PL spectrum of the crystal grown from the Lu$_{2}$Si$_{0.98}$O$_{4.96}$ melt. This fact proves that the PL of the studied crystal at 405 nm (Fig. 1) is extrinsic. Indeed, in the spectrum of the LSO ceramics this band is absent. 

In the spectrum of glassy SiO$_{2}$, there is a band in the region of 350-400~nm with the decay times of several microseconds corresponding to the spin-forbidden transitions in oxygen vacancies~\cite{30,31}. We suppose that the band at 385~nm in the LSO ceramics is analogously related to the spin-forbidden transition in LSO oxygen vacancies. This leads to a significant decrease in the transition probability and, consequently, to an increase in the PL duration.

\begin{table*}
\caption{\label{tab:table3}Kinetic parameters of LSO emission in the region of 250-300 nm}
\begin{tabular}{lllllllll}
Material&$\lambda$, nm&$\tau_{1}$, ns&$a_{1}$, \%&$\tau_{2}$, ns&$a_{2}$, \%&$\tau_{3}$, ns&$a_{3}$, \%&Detalis\\
\hline
Ca$_{0.08}$Lu$_{1.92}$SiO$_{4.96}$&283&11.5$\pm$0.1&53$\pm$1
&28.7$\pm$0.4&47$\pm$1&&&room temperature\\
(ceramics)&&&&&&&&(this work)\\
\\
Lu$_{2}$Si$_{0.98}$O$_{4.96}$&290&5.3$\pm$0.2&25$\pm$2&17.6$\pm$0.3&75$\pm$2&&&room temperature\\
(melt composition,&&&&&&&&(this work)\\
crystal)&&&&&&&&\\
Lu$_{2}$SiO$_{5}$&250&5.2&43&12.2&55&$>$1000&2&temperature of 8K\\
(stoichiometric&&&&&&&&STE emission\\
crystal)&&&&&&&&(Ref.~\cite{10})\\
\end{tabular}
\end{table*}

It should be noted that the spectral and kinetic parameters of studied PL in the region of 280-290~nm are close to the parameters of STE luminescence from stoichiometric LSO without additional doping at liquid helium temperature~\cite{8,9,10}. Indeed, this luminescence is
composed of two bands at $\sim$250 and $\sim$360~nm~\cite{8,9,10}. According to the results of~\cite{10}, the PL kinetics at 250~nm excited by a synchrotron radiation is described by a sum of three exponents:

\begin{equation}
	I(t) = a_{1} \exp \left( -t/\tau_{1} \right) + a_{2} \exp \left( -t/\tau_{2} \right) + a_{3} \exp \left( -t/\tau_{3} \right).
\end{equation}

Fitting by Eq.~(4) made in Ref.~\cite{10} gives $\tau_{1}$~=~5.2~ns, $a_{1}$~=~43\%, $\tau_{2}$~=~12.2~ns, $a_{2}$~=~55\%, $\tau_{3} >$1~$\mathrm{\mu}$s and $a_{3}$~=~2\%. Since the contribution of the third term is small, it can be ignored.
The decay times $\tau_{1}$ and $\tau_{2}$ are very close to the decay times of PL at 290 nm~from the crystal grown from the non-stoichiometric melt Lu$_{2}$Si$_{0.98}$O$_{4.96}$. Comparison of the kinetic parameters for LSO luminescence in the region of 250-300 nm is presented in Table~2.

Synchrotron measurements of decay times are limited by several hundreds of nanoseconds due to the orbital period of electrons. Thereby, due to this limitation, the decay time of the band at 360~nm did not obtain exactly in Ref.~\cite{10}. However, it was observed that the duration of this band exceeds 1~$\mathrm{\mu}$s. It was concluded in Ref.~\cite{10} that the band at 360~nm is related to decay of the STE triplet state and the band at 250~nm is related to the singlet STE decay.

Consequently, the spectra and kinetic parameters of observed PL bands at room temperature from the LSO crystal and ceramics with additional oxygen vacancies in the regions of 280-290 and 385~nm are in general close to the spectral and kinetic parameters of STE luminescence from the stoichiometric LSO at liquid helium temperature. Simultaneously, there is a slight long-wavelength shift in the case of the non-stoichiometric LSO. However, at room temperature, the STE luminescence is completely quenched~\cite{4,8,9,10} in contrast to the luminescence observed in this work.

Nowadays, the STE structure in LSO is unknown. A hypothetical STE model was proposed in Ref.~\cite{4}. According to this model, an STE is localized near the Lu-O bond. The most used conception describing the STE formation in oxides is a bond cleavage between a cation and an oxygen atom. At the next step of the process, a release of oxygen atom into interstitial space occurs and a short-lived Frenkel defect is formed~\cite{32,33}. As a result, a short-lived oxygen vacancy is formed, and its emission is observed.

From this point of view, the following explanation of the phenomena observed in this work can be given. At the temperature higher than 50K when STE is quenched~\cite{4}, a Frenkel defect “oxygen vacancy – interstitial oxygen atom” returns to initial position due to thermal motion, which leads to the thermal destruction of STE before its radiative decay. Simultaneously, oxygen vacancies created in LSO both by growing the crystal from a non-stoichiometric melt and by doping with Ca$^{2+}$ ions are stable at temperatures up to room temperature. The results presented in this work show that the oxygen vacancies luminescence and STE luminescence have close parameters. We suppose that this fact manifests that the structures of STE and stable oxygen vacancies in LSO created by divalent doping and growing of a crystal from oxygen-defficient melt are close to each other.

In Ref.~\cite{4} it was shown that an energy transfer from STE is an important excitation channel of the radiative Ce$^{3+}$ states in LSO. Consequently, based on the similarity in the structures of STE and oxygen vacancies in LSO it can be supposed that the oxygen vacancies stabilized by the Ca$^{2+}$ ions at room temperature can play a significant role in the excitation of radiative Ce$^{3+}$ states in LSO:Ce:Ca crystal analogous to energy transfer from STE. This can lead to the observed increase in scintillation parameters of this composition. Further research is required to verify these hypotheses.

\section{Conclusion}\label{Conclusion}

A growing of non-stoichiometric LSO crystal with the oxygen deficiency and a synthesis of the LSO ceramics doped with Ca$^{2+}$ ions having $\sim$0.5~at.\% oxygen vacancies in its structure lead to formation of a new luminescence band at 280-290~nm. This band is fairly bright at room temperature and the origin of this band is related to oxygen vacancies. 

The nanosecond duration of this luminescence shows that this emission is related to a dipole-resolved optical transition. Also, in the spectrum of LSO ceramics there is a band at 385~nm with the decay time of 15~$\mathrm{\mu}$s which is related to the forbidden optical transition. The spectral and kinetic properties of the observed luminescence bands are close to the spectral and kinetic properties of STE luminescence in stoichiometric LSO at cryogenic temperatures. This points out a similarity between the structures of STE and oxygen vacancies in LSO.

\section{Acknowledgment}
The work is supported by Russian Science Foundation (project \# 19-79-30086)

\end{document}